\documentstyle[twocolumn,aps,psfig,floats]{revtex}

\begin{document}

\def\be{\begin{equation}}
\def\ee{\end{equation}}

\draft

\twocolumn[\hsize\textwidth\columnwidth\hsize\csname %
@twocolumnfalse\endcsname

\title{Phase-Field Model of Mode III Dynamic Fracture}

\author{Alain Karma$^1$, David A. Kessler$^2$, and Herbert Levine$^3$}
\address{$^1$Department of Physics, Northeastern University, Boston MA 02215 \\
$^2$Department of Physics, Bar-Ilan University, Ramat-Gan, Israel\\
$^3$Department of Physics, University of California,
San Diego, La Jolla, CA 92093-0319}

\date{\today}

\maketitle

\begin{abstract}
We introduce a phenomenological continuum model for mode III dynamic fracture that
is based on the phase-field methodology used extensively to model
interfacial pattern formation. We couple a scalar field, which distinguishes
between ``broken'' and ``unbroken'' states of the system, to
the displacement field in a way that consistently includes
both macroscopic elasticity and a simple rotationally invariant
short scale description of breaking.
We report two-dimensional simulations that yield
steady-state crack motion in a strip geometry above the Griffith threshold.
\end{abstract}
]

The nonequilibrium physics of dynamic fracture continues to be a
challenging topic of great interest \cite{review}.
Recent efforts have been spurred by
experimental findings regarding the breakdown of straight crack
propagation (along with its associated smooth fracture surface) as
the crack speed exceeds a threshold value. This instability has
been seen in a variety of materials, both
crystalline~\cite{cramer} and amorphous~\cite{exp1,exp3}, and it has
been reproduced in molecular dynamics
simulations~\cite{mdsim} as well as with finite element
schemes~\cite{finite}.

For any material undergoing brittle fracture, linear continuum
elasticity only provides an accurate description of the
displacements in regions that are not too close to the crack tip.
The classic approach to this problem~\cite{review} has been to
solve linear elastic equations, with boundary conditions
providing the driving stresses, right up to this tip. This approach
relies upon the assumption that the ``process zone" inside of
which the linear continuum equations break down is microscopic in
size. The solutions have stress fields which become singular at
the assumed tips, representing a finite flow of energy into the
infinitesimally sized process zone. The velocity of the crack is
then phenomenologically assumed to be given by some function of
this energy flow rate.

This approach has two main limitations from a physics perspective.
Firstly, it does not provide insight into how the crack velocity is
actually determined, e.g. how it depends on short-scale
dissipation. Secondly, and more importantly, it fails to
predict instabilities of the tip dynamics.
Thus, just as is the case in the well-studied
problem of dendritic solidification~\cite{kkl},
one must supplement the macroscopic transport physics with a
consistent, regularizing microscopic theory on the tip scale in
order to create a sensible theoretical framework.

One method for accomplishing this task involves placing the system
on a lattice and allowing for the elastic forces to rapidly
diminish at large atomic separation.
Analytical~\cite{slepyan,mg,pre-old,leonid} and
numerical~\cite{pre-old,pre-new} studies of such models have shown
that the details of the lattice structure are critical for the tip
dynamics. This is not surprising since in general the process zone
scale is just the lattice spacing. Thus, these models are useful
but cannot even qualitatively describe experiments in amorphous
systems. What appears to be a more sensible approach for this
class of systems is to construct a regularized continuum model
that maintains {\em rotational symmetry} even inside the process
zone. Constructing such a theory is the aim of this
paper~\cite{cohesive}.

As a first step, we focus here on
the simpler situation of mode III fracture for
which the displacement $\vec{u}$ can be taken to be in a fixed
direction (out of the plane) and hence can be represented by a
scalar field $u$. Standard linear elasticity assigns the energy
\begin{equation}
E\ = \ \int d\vec{x} \ \frac{1}{2}\, \mu\, \vec{\epsilon} ^{\ 2}
\label{linear-energy} \end{equation} with the strain
$\vec{\epsilon} = \vec{\nabla} u$ and an elastic constant $\mu$.
Allowing a  material to fracture means that at large enough
$\vec{\epsilon} ^{\, 2}$, the energy becomes strain independent,
thereby eliminating the force. In an ideally brittle material, for
example, this transition occurs immediately at some critical
magnitude of the strain, $\epsilon _c$. Our basic idea involves
representing the local state of the system, either unbroken with
$|\epsilon | < \epsilon _c$ or broken with $|\epsilon | > \epsilon
_c$, via a second ``phase" field $\phi (\vec{x},t)$. This field
can be made to track the correct state if it obeys a standard
two-minimum Ginzburg-Landau equation with the relative energy of
the two wells dependent on $\epsilon ^2 - \epsilon _c^2$.
Specifically, we choose
\begin{equation}
\tau \,\partial_t \phi (\vec{x},t)   =  D_{\phi}
\nabla^2 \phi  -  V'_{DW} (\phi  ) - \frac{\mu}{2} \,
g'(\phi ) \left( \vec{\epsilon} ^{\ 2} - \epsilon _c^2 \right)
\label{phi-equation}
\end{equation}
where $V_{DW}(\phi)=\frac{1}{4}\phi^2 (1-\phi )^2$ and $g$ is a
function specified later that has the
properties $g(0)=0$, $g(1)=1$ and $g'(0)=g'(1)=0$. With
these choices, the two minima are always at $\phi =0$ and $\phi
=1$, with the absolute minimum shifting from 1 to 0 as
$\vec{\epsilon}^{\ 2}$ passes $\epsilon _c^2$.

To close the system, we need to specify how $\phi$ affects the
elasticity equation. Note that the above equation follows from
the relaxational dynamics, $\tau\partial_t\phi=-\delta E/\delta\phi$,
where the energy $E$ is now given by
\begin{equation}
\int d\vec{x} \left[ \frac{1}{2} D_{\phi} (\vec{\nabla} \phi )^2
+V_{DW} (\phi ) +\frac{\mu}{2}\, g(\phi ) \left( \vec{\epsilon} ^{\
2} - \epsilon _c^2 \right) \right] \label{phi-energy}
\end{equation}
If we now interpret this $E$ as
the full potential energy including the elastic contribution, we
see that the aforementioned properties of $g(\phi )$ will in fact
eliminate the elastic force for large strain without having any
effect at small strain where $\phi \simeq 1$ and thus $g \simeq
1$. Consequently, our second equation is derived by varying this energy
with respect to displacement, which yields
\begin{equation}
\rho \frac{\partial ^2 u}{\partial t^2} + b \frac{\partial
u}{\partial t} \ = \ \mu \vec{\nabla} \cdot \left( g(\phi )
\vec{\nabla}  \left( 1+ \eta \frac{\partial}{\partial t} \right) u
\right), \label{u-equation}
\end{equation}
where we have allowed for both a Stokes drag term with coefficient
$b$ and a Kelvin viscosity $\eta$. This equation completes our
model specification.

Let us place this work in some perspective. Our approach is
similar in philosophy to, but very different in detail from, the work of
Aranson and co-workers~\cite{aranson} who also derive a continuum
two-field model for fracture. Most crucially, the physical
interpretation of the phase field and hence the way in which it
enters dynamically are completely different. As a result the present model
avoids certain unphysical features of their model, e.g. the logarithmic
dependence of the crack opening on the system size.
Within the traditional fracture community, several
researchers~\cite{fourth-order} have studied the effects of
``softening" the elastic energy at large strain and compensating
for the resultant instability in the equations by adding higher
derivative terms. This approach, however, turns out to be
difficult to extend to construct a continuum model
where the fracture energy is independent of the system size and,
at the same time, the strain is fully relieved in the bulk solid
after passage of the crack. Moreover, it leads to a single
fourth-order elasticity equation that is extremely hard to treat
numerically. Finally, it is
worth recalling that the original idea~\cite{collins} of
representing different phases of a system via a field coupled to
the macroscopic dynamics, and thereafter using derivative terms in
the phase field to regularize the problem, arose in the context of
nonequilibrium crystal growth where it has become the method of
choice\cite{karma+rappel,goldenfeld} for highly accurate
computations of solidification microstructures.

To understand how our model accounts for fracture, we start with
the (one dimensional) snap-back of a stretched elastic band of
size $2L$ after it breaks in the middle. Let us first consider the
final time-independent cracked state; note that this is
equivalently the asymptotic state for a 2-d crack once the tip has
passed. This state is determined by solving the above equations
with all time derivatives set to zero, with the boundary
conditions $u(\pm L)= \pm \Delta$, $\phi(\pm L)=1$. Note that
$2\Delta$ is the total integrated strain that is conserved by
the dynamics. Moreover, both $\epsilon(y)$ and $\phi(y)$ are symmetrical
about $y=0$; thus we only need to find a
solution in the interval $0\le y \le L$.
The elasticity equation, Eq. \ref{u-equation},
requires $ \epsilon (y) = \partial_y u= \epsilon _0 /g(\phi
(y))$. Substituting this form into Eq. \ref{phi-equation} yields
\begin{equation}
0 \ = \ \ D_{\phi} \phi '' \, - \, V'_{DW} (\phi  )\, - \,
\frac{\mu}{2} g'(\phi ) \left( \frac{\epsilon_0 ^2}{g ^2(\phi )} -
\epsilon _c^2 \right) \label{ss-equation}
\end{equation}
From now on, we rescale lengths to make $D_\phi =1$. Eq.
\ref{ss-equation} can be thought of as the equation of motion of
a ball rolling in an effective potential
\begin{equation}
V _{EFF} (\phi ) \ =\ -V_{DW} +\frac{\mu}{2} \left( g(\phi )
\epsilon _c ^2 + \frac{\epsilon_0^2}{g(\phi )} \right)
\label{v-effective}
\end{equation}
A schematic picture of this potential is shown in Fig. 1. The
solution of interest corresponds to rolling in a ``time" $L$ from
the top of the hill at $\phi =1$ to the turning point $\phi ^*$
located near $\phi =0$; this turning point exists because
$ \epsilon_0^2/g(\phi )$ becomes large and positive for
small $\phi$.
\begin{figure}
\centerline{ \psfig{file=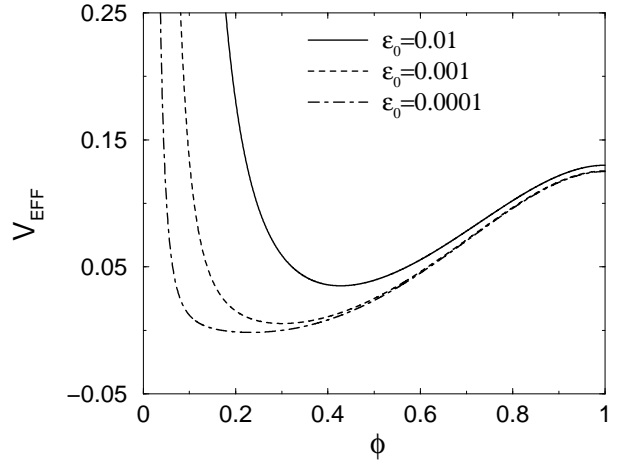,width=.4\textwidth}}
\smallskip
\caption{Plots of the effective potential for 1-d static crack
profiles ($\mu=1$ and $\epsilon_c=1/2$).}
\label{fig1}
\end{figure}

The asymptotic steady-state crack is thus given by
\begin{equation}
\frac{1}{\sqrt{2}}\,
\int_{\phi ^*}^{\phi(y)} \ \frac{d \tilde{\phi}}{\sqrt{E_0 - V_{EFF}
(\tilde{\phi})}} \ = \ y \label{crack-solution} \end{equation} The
two unknown constants $\epsilon_0$ and $E_0$ are fixed by the
requirements that the above equation yields $\phi = 1$ at $y=L$
and by the overall integrated strain constraint
\begin{equation}
\frac{\epsilon_0}{\sqrt{2}}\,\int_{\phi ^*}^{1} \ \frac{\, d \tilde{\phi}}
{g(\tilde{\phi} ) \sqrt{E_0 - V_{EFF} (\tilde{\phi})}} \ = \
\Delta \label{strain}
\end{equation}
For this solution to be physically acceptable, almost all of the displacement
must occur in the crack, thereby relieving the strain in the bulk,
which imposes a constraint on the form of the function $g$.
To see why, consider the large $L$ limit (where with $D_\phi =1$ our length
unit is the process zone scale) and note that the standard
Griffith criterion for fracture necessitates the scaling $\Delta
\sim \sqrt{L}$. There is a contribution to the integral on the
left hand side of Eq. \ref{strain} that arises from $\phi$ of
order $\phi ^*$, which is close to zero. If we choose $g$ to vanish
near $\phi =0$ as a power law $g \sim \phi^{2+\alpha}$, it is
clear from the form of the effective potential that $ \phi ^* \
\sim \epsilon_0 ^{\frac{2}{2+\alpha}}$. Given this, the local
contribution to the integral scales like $\epsilon_0
^{-\frac{\alpha}{2+\alpha}}$. This will match the right hand side
with the choice
\begin{equation}
\epsilon_0 \ \sim \ L^{-\frac{2+\alpha}{2\alpha}} \nonumber
\end{equation}
Hence, as long as $\alpha$ is positive, $\epsilon_0$ will go to
zero at large $L$ fast enough such that the local contribution,
i.e. that of the crack, to the overall displacement is dominant
compared to the bulk contribution which scales as $\epsilon_0 L$.
We note that the least residual stress occurs in the limit $\alpha
\rightarrow 0^+$, which gives rise to an exponential decrease as a
function of system size. Finally, the
fracture energy $\gamma$ remains finite as $L$ gets large, as is also
required for a sensible theory. In this limit, $\epsilon_0\rightarrow 0$
and $E_0\rightarrow \mu\,\epsilon_c^2/2$, and it is easy to derive
the expression
\begin{equation}
 \gamma =\sqrt{2} \int_0^1 d\tilde\phi
 \sqrt{\mu\,\epsilon_c^2/2-V_{EFF}(\tilde\phi;\epsilon_0=0)}\label{gamma}
\end{equation}
We now turn to the 1d time-dependent problem. We choose
$g=4\phi^3 - 3\phi^4$, so that $\alpha=1$ and $\epsilon_0$
should
\begin{figure}
\centerline{ \psfig{file=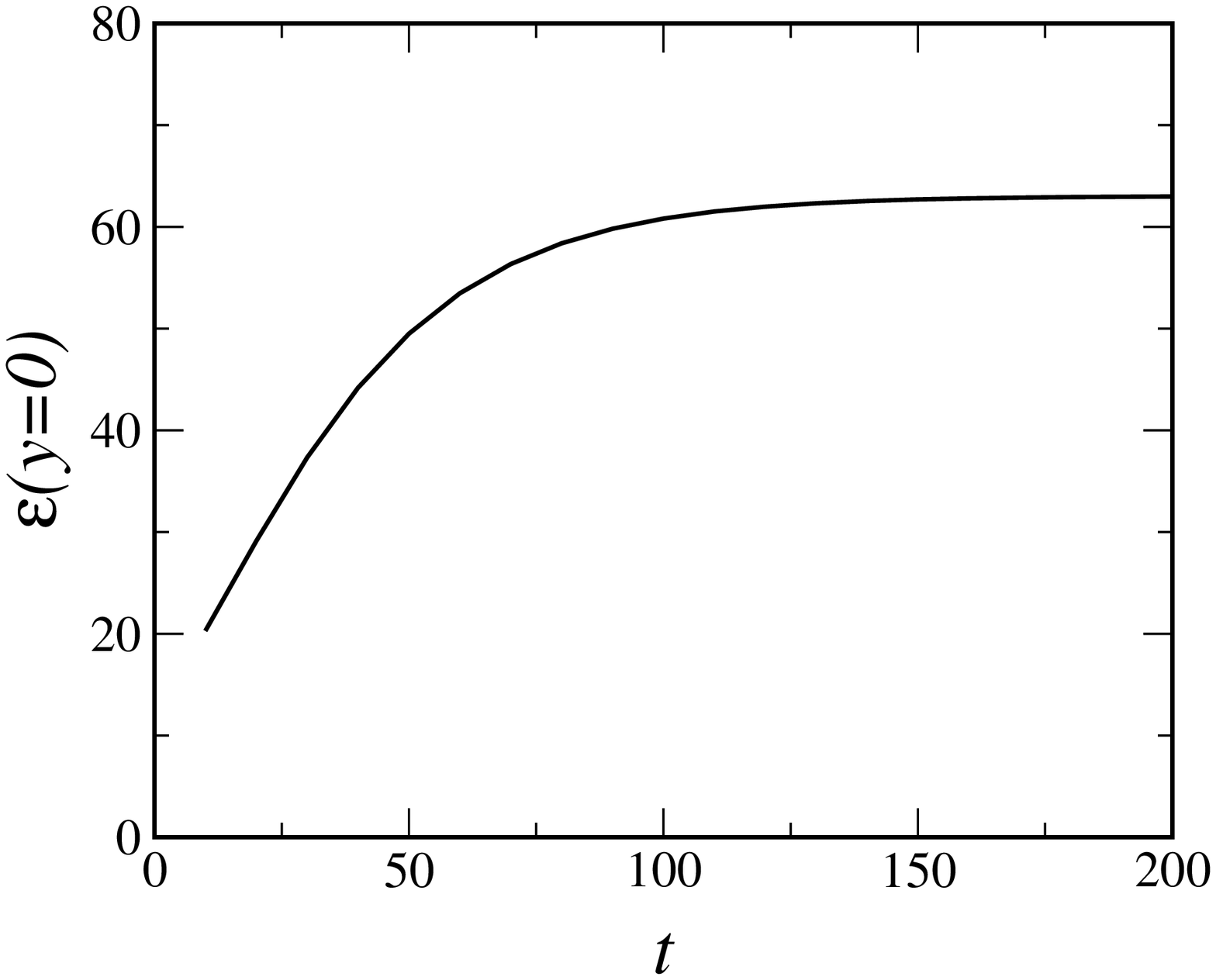,width=.35\textwidth}}
\centerline{ \psfig{file=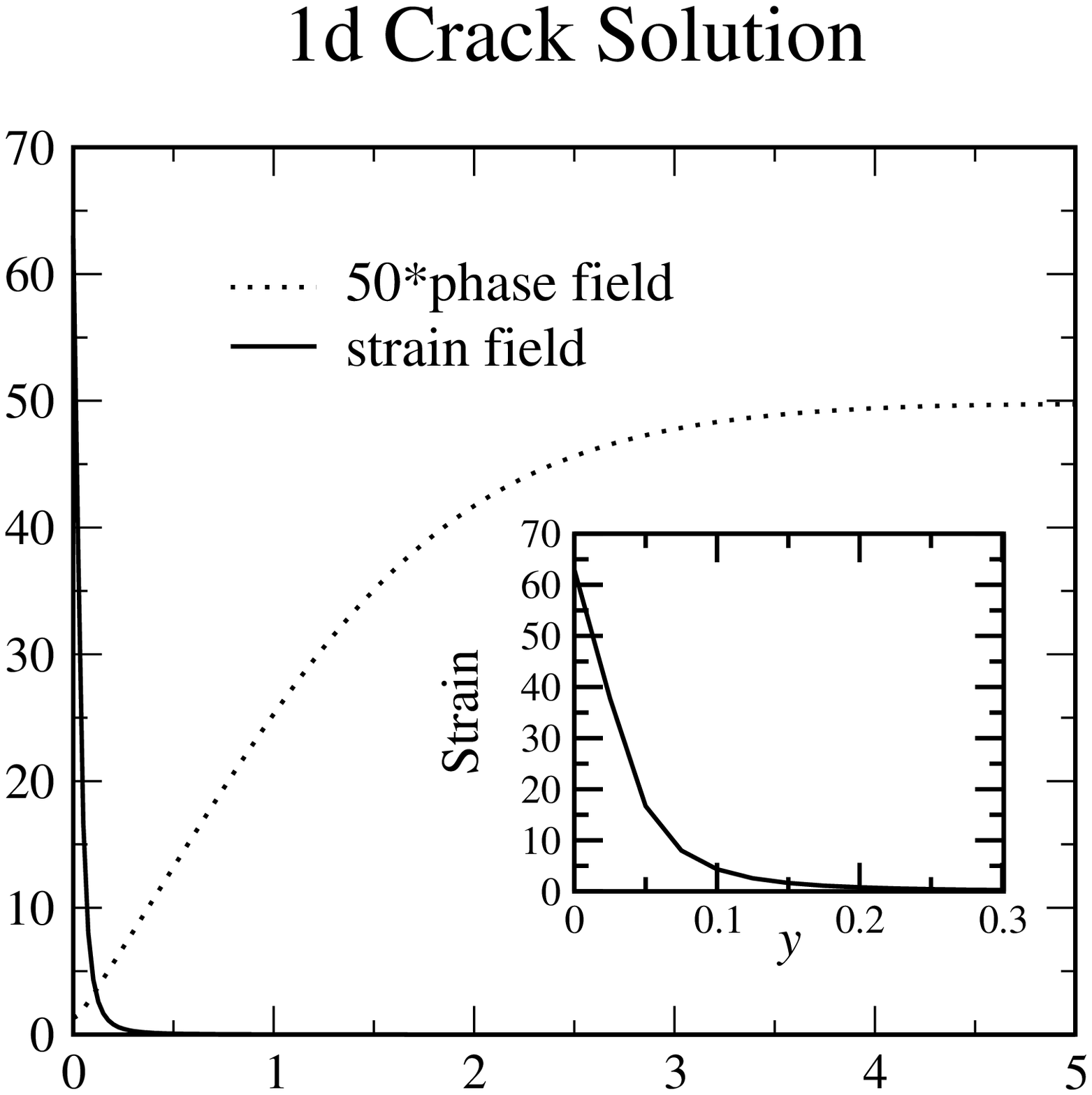,width=.32\textwidth}}
\smallskip
\caption{Results of 1d crack computations. (a) The
strain at the crack center vs. time (b) asymptotic profiles of the
strain and phase fields. $L=5$, $\epsilon_0=0.045$
}
\label{fig2}
\end{figure}
\noindent
scale as $L^{-3/2}$; also we pick $\epsilon _c =1/2$, $\mu =1$. Fig. 2 shows the time development of the
strain at the origin, for a completely overdamped system,
$\rho=0$, $b=1$, $\eta=0$.  We see that, starting from an initial
large strain region near the origin,
 the system proceeds to fully crack
and approach the aforementioned asymptotic state. Notice that for
this case of overdamped dynamics, there are two time scales
visible in the strain relaxation; a fast time scale during which
the crack (say in the $\phi$ field) develops and a slower one
during which strain is drained from the bulk into the crack. For a
typical underdamped case, ($\rho=1$, $b=0$, $\eta =0.2$), the
second-stage relaxation to the steady-value involves, as it must,
damped oscillations. In either case, in the second stage, the
$\phi$ field is slaved to the displacement making the system of
equations very stiff. We coped with this difficulty using an
implicit time-stepping scheme.

The real test of any fracture model comes in two (or higher)
dimensions. Now, the crack tip must advance by providing enough
stress to strain new material beyond a critical extension.
We have carried out a preliminary simulation study of our model in
a 2-d strip geometry (with the edges of the strip
at $y=\pm L$) using a standard Crank-Nicholson
alterning-direction-implicit scheme ~\cite{numrec}. We used
the initial condition corresponding to a strained solid with $\phi(x,y)=1$ and
$u(x,y)=\Delta y/L$, which must produce a crack propagating along the $x$-direction
above the standard Griffith threshold, $\Delta >\Delta_c$, where
here $\Delta_c=\sqrt{2\gamma L}$ and $\gamma=0.3808...$ is given by
Eq. \ref{gamma} for the present model parameters
($\mu=1$, $\epsilon_c=1/2$).
A typical time sequence for a
stably propagating crack is presented in Fig. 3. We have checked
that our results are reasonably independent of the discretization scale $dx$,
such that we are truly seeing the results of the continuum regularization of
the tip scale dynamics.

In Fig. 4, we present the steady-state crack velocity as a
function of the driving. The crack propagates above
a critical value of the drive that is within a few percent of the
analytically predicted value $\Delta_c=\sqrt{2\gamma L}$ for
the large $L$ limit even though $L$ is not so large ($L=10$) in
the simulations. It is important to recognize that $V(\Delta)$
cannot be obtained from the usual continuum theory
without additional assumptions; here it follows directly from the
fact that we have a consistent theory at both the macroscopic and
microscopic scales. One can obtain similar results from lattice
models of fracture~\cite{mg,pre-old,new-slepyan}, at the price of
introducing lattice scale instabilities~\cite{marder-liu,pre-new}.
These instabilities are connected to spatial period-doubling, when
the times for breaking alternating diagonal bonds in a hexagonal
lattice become unequal. Thus, they have no direct relevance for
amorphous systems without any underlying crystallinity. Here, at
least at the drivings we have investigated so far, we find no such
instability, and steady cracking appears to persist to rather high
displacements. In terms of experiment, this seems to imply that the
observed transition at moderate displacement to more complex tip
dynamics is presumably connected to the mode I nature of the
fracture geometry, with its propensity to branch in the direction
of the off-axis stress maximum as suggested originally by Yoffe~\cite{yoffe}. An extension of the present approach
to mode I is presently underway to test this hypothesis.

In terms of physics, our approach leads to the introduction of a
new time scale, $\tau$, connected to the relaxational rate of the
phase-field. At any non-zero $\tau$, this relaxation is a possible
source of tip-scale dissipation and hence it can affect the crack
propagation. Decreasing this parameter in the simulations
indeed increases the velocity, consistent with convergence to
a finite limit as $\tau\to 0$ with $O(\tau)$ corrections. This
indicates that the velocity is predominantly limited here
by the rate at which the strain is drained
from the bulk into the crack.

An important numerical issue for future consideration concerns the
sharpness of the strain profile. To fully resolve the spatial
scales in the 2-d crack is a daunting task which probably cannot be
accomplished by sticking with a fixed computational grid. A
related issue concerns the long time scale necessary for full
strain relaxation. It may be possible to modify the time
derivative terms in the equation of motion to speed up this relaxation.
Furthermore, to make contact with
expected results from the fracture community (such as the role of
the stress intensity factor and the fact that crack tips will
propagate without inertia at least until times such that shear
wave reflections from the boundaries can affect the motion) will
require much larger systems and much more attention to details of
the initial conditions.
\begin{figure}
\centerline{ \psfig{file=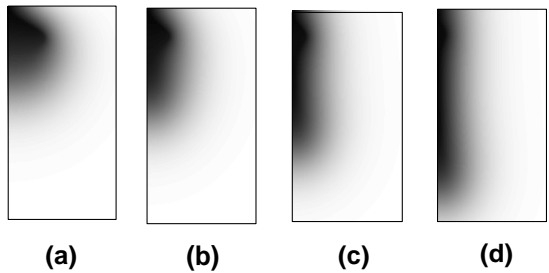,width=.4\textwidth}}
\smallskip
\caption{2-d simulation snapshots; pictured
here is $\phi$ in grey-scale (0=black, 1=white) over
1/2 the computational domain of
size 20x20; here $\rho=1$, $b=0$, $\eta=.2$, $\Delta = 2.81$, timestep=.02 and grid spacing =
.05}
\end{figure}
\begin{figure}
\centerline{ \psfig{file=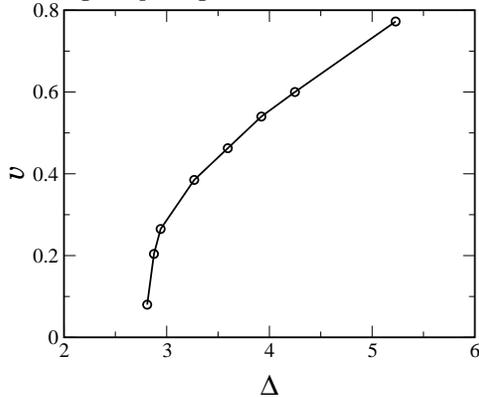,width=.35\textwidth}}
\smallskip
\caption{Velocity vs. driving from 2-d simulations.}
\label{fig4}
\end{figure}


The authors thank the hospitality of the Aspen Center for Physics, where
this work was begun. DAK acknowledges the support of the Israel Science
Foundation. AK acknowledges support of DOE and thanks Jim Rice
for valuable discussions in the initial stage of this work.


\begin{thebibliography}{99}
\bibitem{cohesive} Attempts to do this by merely modifying the boundary conditions
(so-called cohesive zone models~\cite{barenblatt+langer}) have not
been successful to date~\cite{langer-lobkovsky}.
\end{thebibliography}
\end{document}